\documentclass[12pt,twoside]{ord}
\usepackage[T1]{fontenc}
\usepackage{amsfonts,amsmath,bm,bbm}
\usepackage{graphicx}
\usepackage{epstopdf}
\usepackage{amssymb}
\usepackage{mathtools}
\usepackage{enumerate}
\usepackage{natbib}
\usepackage{indentfirst}
\usepackage[font=footnotesize]{caption}
\usepackage{floatrow}
\usepackage{enumitem}
\DeclareMathOperator{\argmin}{argmin}

\usepackage{url}
\usepackage{tikz}
\usepackage{varwidth}
\usetikzlibrary{patterns,decorations.pathreplacing}
\usepackage{booktabs}
\usepackage[normalem]{ulem}
\useunder{\uline}{\ul}{}
\def\b#1{\beta_{#1}}

\usepackage{color}
\definecolor{gray}{rgb}{0.5,0.5,0.5}
\definecolor{dgreen}{rgb}{0,0.5,0}

\ORDtitle{Regularization for electricity price forecasting}
\ORDshortafill{B. Uniejewski}
\ORDshorttitle{Regularization for electricity price forecasting}
\ORDaffilone{Department of Operations Research and Business Intelligence, Wroclaw University of Science and Technology, Wroclaw, Poland}
\ORDaffiltwo{}
\ORDaffilthree{}
\ORDaffilfour{}
\ORDauthorone{Bartosz~Uniejewski}{*1}{0000-0001-8563-2514}{}
\ORDauthortwo{}{}{}{}
\ORDauthorthree{}{}{}{}
\ORDauthorfour{}{}{}{}
\ORDauthorfive{}{}{}{}
\ORDcorrespoding{bartosz.uniejewski@pwr.edu.pl}

\ORDnumber{xx}
\ORDvolume{x}
\ORDdoi{DOI: draft}
\ORDyear{xxxx}
\ORDReceived{xxxxx~xxxxx}
\ORDRevised{xxxxx~xxxxx}
\ORDAccepted{xxxxx~xxxxx}
\ORDPublished{xxxxx~xxxxx}
\ORDFirstpagenum{1}

\begin{document}

\maketitle

\begin{abstract}
	The most commonly used form of regularization typically involves defining the penalty function as a $\ell^1$ or $\ell^2$ norm. However, numerous alternative approaches remain untested in practical applications. In this study, we apply ten different penalty functions to predict electricity prices and evaluate their performance under two different model structures and in two distinct electricity markets. The study reveals that LQ and elastic net consistently produce more accurate forecasts compared to other regularization types.  In particular, they were the only types of penalty functions that consistently produced more accurate forecasts than the most commonly used LASSO. Furthermore, the results suggest that cross-validation outperforms Bayesian information criteria for parameter optimization, and performs as well as models with \emph{ex-post} parameter selection.
\end{abstract}


\ORDKeywords{electricity price forecasting, regularization, power market, convex regularization, LQ regularization, elastic net}

\section{Introduction}

\emph{Electricity price forecasting} (EPF) is a critical task for many participants in the power market, including power generators, traders and consumers. Accurate predictions of electricity prices enable to make informed decisions regarding production, purchasing, and sales \citep{nar:zie:22, nit:wer:23}. In particular, forecasting of day-ahead market prices is of great importance for decision-making in electricity markets \citep{mac:22}. This auction-based market plays a crucial role in ensuring the reliability and efficiency of the power system \citep{may:tru:18}. The day-ahead market not only gives the opportunity to adjust the long-term position to actual exposure \citep{kat:zie:18,jan:woj:22}, but is also a reference point for \textit{over-the-counter} (OTC) trading and settlement.

One of the main challenges in developing reliable electricity price forecasting models is the high volatility of the electricity market \citep{wer:14}. Many traditional forecasting models fail to capture these complex dynamics, resulting in inaccurate forecasts \citep{lag:mar:sch:wer:21}. To overcome this problem, researchers have turned to regularization techniques to improve the accuracy of their models. Regularization is a method used in machine learning and statistical modeling to prevent overfitting of a model by adding a penalty term to the model's objective function.

Although regularization has been widely applied in machine learning and related fields, its use in electricity price forecasting is still limited. To our best knowledge, up to now, only three out of ten considered in this paper penalty functions have been tested in the EPF context. The first research that incorporated regularization in this context was conducted by Barnes and Balda~\cite{bar:bal:13}, who used ridge regression to assess the financial viability of battery storage in electricity markets. More recently,  Ludwig~et~al.~\cite{lud:feu:neu:15} and Ziel~et~al.~\cite{zie:ste:hus:15} applied the \emph{least absolute shrinkage and selection operator} (LASSO) to improve prediction accuracy. Lastly, Uniejewski~et~al.~\cite{uni:now:wer:16} utilized elastic net to automate variable selection in the electricity market. In this paper, we investigate the performance of ten different types of regularization to forecast electricity prices. The techniques we consider are adaptive LASSO, clipped LASSO, concave potential function, elastic net, forward-LASSO adaptive shrinkage, LASSO, LQ,  minimax concave PLUS, ridge regression and smoothly clipped absolute deviation. 

We assess the performance of those regularization types using two datasets from the German EPEX SPOT and Iberian OMIE markets and two model structures well-established in the literature. Additionally, in this paper, we employ \emph{cross-validation} (CV) or \emph{Bayesian information criteria} (BIC) to select the value of the tuning parameter ($\lambda$) and compare their performance. Finally, we propose a fully automated approach to select all the regularization parameters based on CV. The accuracy of price forecasts obtained with regularized models is compared with the benchmark models estimated with \emph{ordinary least squares} (OLS).

The paper is structured as follows. In Section \ref{sec:data}, we provide a brief overview of the datasets. Next, in Section \ref{sec:methodology} we outline the regularization techniques considered and summarize related research on electricity price forecasting. In Section \ref{sec:results}, we compare the accuracy of the forecasts obtained with the proposed methods, and finally in Section \ref{sec:conclusions} we conclude the results.

\section{Datasets}
	\label{sec:data}

	\begin{figure}[t]
		\centering

  \includegraphics[width=0.85\textwidth]{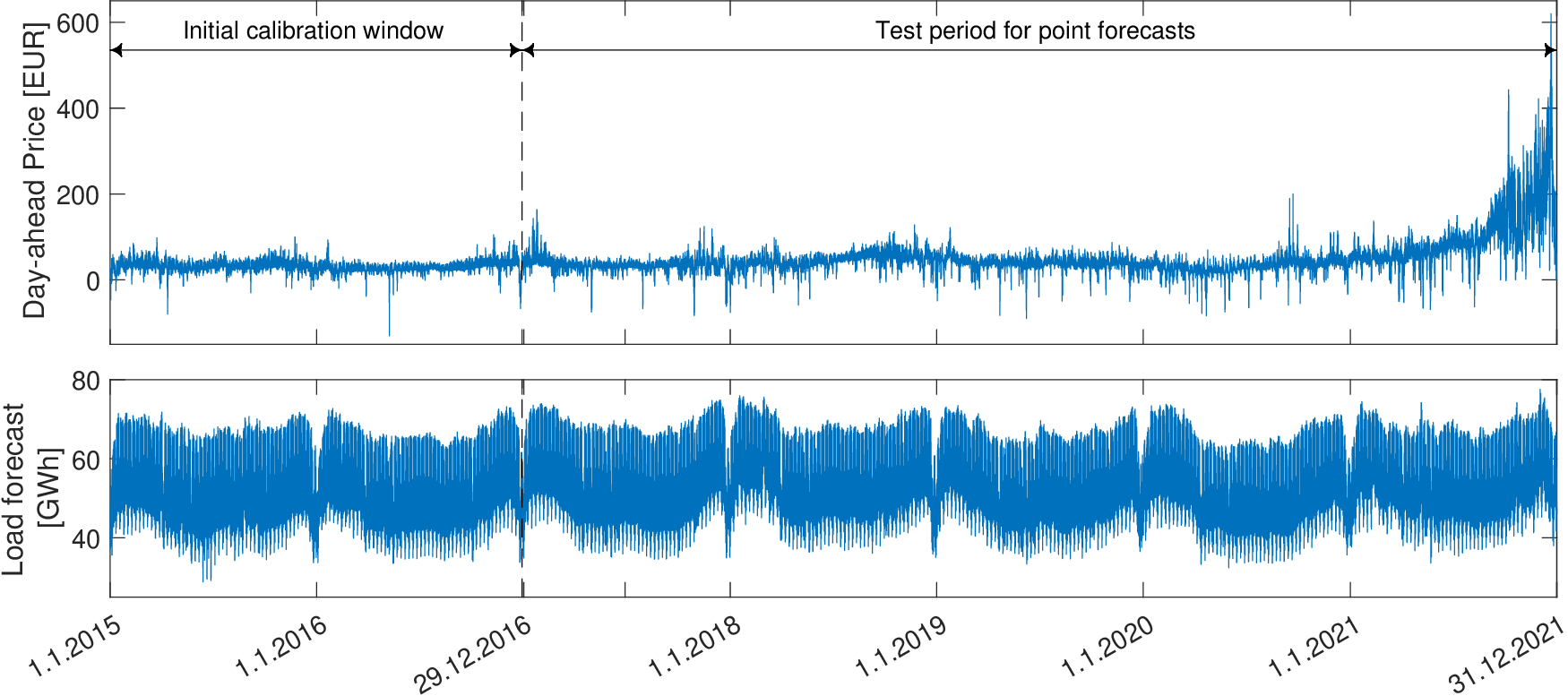}
  \includegraphics[width=0.85\textwidth]{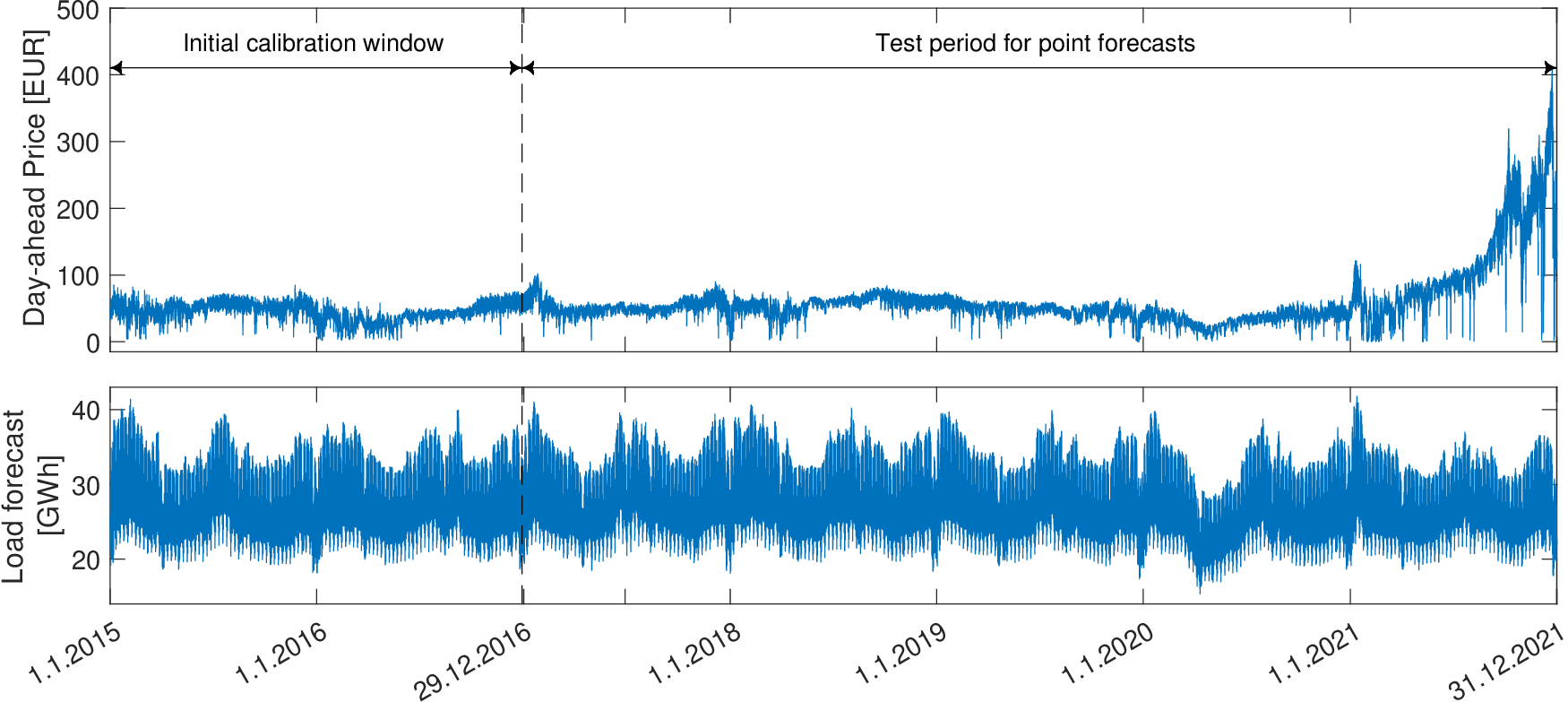}

		\caption{Day-ahead prices and day-ahead load forecast time series for German EPEX (top panel) and Spanish OMIE (bottom panel) from 1.1.2015 to 31.12.2021. The vertical dashed line marks the beginning of the out-of-sample test period (29.12.2016).}
		\label{fig:data1}
	\end{figure}
	
This empirical study utilized datasets from two distinct European markets, each covering a period of seven years. The first market is the German EPEX SPOT. It has recently experienced a significant increase in the share of renewable generation and underwent major structural changes, making it an attractive target for analysis. The data were obtained from the transparency platform (\url{https://transparency.entsoe.eu}) and consist of five hourly time series (see top panel in Figure \ref{fig:data1}):
\begin{itemize} \itemsep 0pt
	\item day-ahead electricity prices for the DE-AT-LU bid zone until 30.9.2018 and DE-LU afterwards
	\item day-ahead total load forecasts for Germany
        \item day-ahead solar generation forecasts for Germany
	\item day-ahead wind (on-shore) generation forecasts for Germany
        \item day-ahead wind (off-shore) generation forecasts for Germany
\end{itemize}

The second market examined in the study is the Spanish OMIE market. Unlike the German market, the Iberian one does not allow prices to drop below zero. This feature, along with the rapid development of the market, makes it an interesting subject for analysis. Data were also collected from the transparency platform and comprise four hourly time series (bottom panel in Figure \ref{fig:data1}).
\begin{itemize} \itemsep 0pt
	\item day-ahead marginal electricity prices for Spain
	\item day-ahead total load forecasts for Spain
        \item day-ahead solar generation forecasts for Spain
	\item day-ahead wind (on-shore) generation forecasts for Spain
\end{itemize}

It should be noted that until 30.06.2021, the Spanish electricity market had a price cap between 0 and 180 Euro/MWh. In the second half of 2021, this limit was changed to a wider price range between -500 and 3 000 Euro/MWh.  However, as the change does not affect the price series, but rather was caused by a change in price dynamics, it has not been included in the model.
	
Both time series span 2557 days, ranging from 1.1.2015 to 31.12.2021. Missing or doubled values (corresponding to the time change) were replaced by the average of the closest observations and the arithmetic mean of the values from doubled hours, respectively.

The structure of the price series in both European markets underwent notable changes during the seven-year period under study, as shown in Figure \ref{fig:data1}. The onset of 2020, coupled with the emergence of the COVID-19 pandemic, followed by the Russian invasion of Ukraine in 2022, induced substantial shifts in electricity price dynamics in 2020-2021.
 
\section{Methodology}
\label{sec:methodology}

The volatility and irregularity of electricity prices pose a challenge in producing accurate forecasts. Outliers in the data, caused by sudden spikes, can distort model coefficients and lead to higher in-sample errors for non-spiky periods. To address this issue, Uniejewski~et~al.~\cite{uni:wer:zie:18} have proposed various functions of \emph{variance stabilizing transformations} (VST) to reduce the variation in the data and improve forecast accuracy. In this paper, the \emph{normal probability integral transform} (N-PIT) is applied to each (price and exogenous variables) time series: $$x_{d,h} = N^{-1}(\hat F_{X_{d,h}}(X_{d,h})),$$ where $\hat F_{X_{d,h}}$ is a empirical cumulative distribution function of sample $X_{d,h}$ and $N^{-1}$ is the inverse function of the standard normal distribution. 

In the next step, the transformed time series are used as the inputs of the forecasting models which are estimated via OLS or one of ten different regularization types. After generating forecasts of the transformed price ($\hat p_{d,h})$, the inverse transformation is applied to obtain the final forecasts of the electricity price $\hat{P}_{d,h} = \hat F^{-1}_{P_{d,h}}(N(\hat p_{d,h}))$.

\subsection{Models}

The price forecasting task is implemented separately for each hour. For each day of the out-of-sample period, this results in 24 different parameter sets. This so-called multivariate framework is a common approach in the EPF literature (see \cite{zie:wer:18} for a discussion of the differences between the uni- and multivariate frameworks). The 24 individual models are independent of each other, but their estimation is based on the same set of information.

The day-ahead forecasts of the hourly electricity price for both markets are determined within a rolling window scheme using a 728-day calibration window. First, all considered models are calibrated to data from the initial calibration period (1.1.2015 - 29.12.2016), and forecasts for all 24 hours of the next day (30.12.2016) are determined. Then, the window is rolled forward by one day; the models are re-estimated, and forecasts for all 24 hours of 31.12.2016 are computed. This procedure is repeated until the predictions for the 24 hours of the last day in the sample (31.12.2021) are made.

The optimal choice of calibration window length is the subject of lively discussion \citep{hub:mar:wer:19, ser:uni:wer:19}, in this paper the 728-day calibration window is used. The choice was based on limited empirical tests comparing the prediction accuracy of models calibrated to the considered window and three shorter calibration sizes.

\subsubsection{Parsimonious ARX model}

To assess the impact of different types of regularization on prediction accuracy, two different model structures are used in this study. The undoubted advantage of using regularization is the almost unlimited number of initially considered explanatory variables in the model. As a result, expert knowledge becomes less important. For this reason, studies incorporating regularization usually focus on the parameter-rich model structures, and the concept is very rarely considered for estimating models with only a few inputs.

To address this literature gap, as the first underlying model, we propose a parsimonious autoregressive structure, which is a well-established model in the EPF literature \citep{zie:wer:18}. The transformed price on day $d$ and hour $h$ (i.e.,\ $p_{d,h}$) is modeled by the following equation:	

	\begin{align*}
	p_{d,h} & = \b{1} p_{d-1,h} + \b{2} p_{d-2,h} + \b{3} p_{d-7,h} + \b{4} p^{max}_{d-1} + \b{5} p^{min}_{d-1} + \b{6} p_{d-1,24} + \\
 & + \b{7} l_{d,h} + \b{8} s_{d,h} + \b{9} w^{on}_{d,h} + \b{10} w^{off}_{d,h} + \sum_{i=1}^7\b{10+i} D^i_{d} + \varepsilon_{d,h}, 
	\end{align*}

where $p_{d-1,h}$, $p_{d-2,h}$ and $p_{d-7,h}$ account for the autoregressive terms and correspond to prices from the same hour of the previous day, two days before, and a week before, respectively. The midnight price of electricity for the previous day, which is the last known price at the time of prediction, is represented by $p_{d-1,24}$. $p^{max}_{d-1}$ and $p^{min}_{d-1}$ are the maximum and minimum prices of the previous day. The transformed day-ahead load forecast for a given hour of a day is represented by $l_{d,h}$, while transformed day-ahead forecasts of solar, on-shore wind and off-shore wind generation are represented by $s_{d,h}$, $w^{on}_{d,h}$, and $w^{off}_{d,h}$, respectively. Finally, $D_d^{1},..., D_d^{7}$ are weekday dummies and $\varepsilon_{d,h}$ is the noise term. Note that off-shore wind generation is excluded from the model for the OMIE market, as there is no installed capacity in Spain. The model consisting of 17 (16 for OMIE) regressors is referred to as ARX throughout the paper.

\subsubsection{Parameter rich model}

The second model structure, denoted by the fARX model, is a more complex autoregressive model that incorporates 277 (229 for OMIE) regressors. The first 72 regressors ($p_{d-i,h*}$) refer to the autoregressive terms and consist of information on prices from all hours from one, two and seven days before forecasted day $d$. The next six variables account for the non-linear terms and are reflected by the minimum ($p^{min}_{d-i}$) and maximum ($p^{max}_{d-i}$) of all prices from one, two, and three days before day $d$. Another 192 regressors (144 for OMIE) refer to exogenous variables. The model uses information on prediction of load ($l_{d-i,h}$), wind ($w^{on}_{d-i,h}$ and $w^{off}_{d-i,h}$) and solar generation ($s_{d-i,h}$) for all hours of the day $d$ and the previous day. Lastly, similarly to ARX model, to capture the weekly seasonality, the model consists of seven dummies variables $D_d^{1},..., D_d^{7}$. The formula for the fARX model is as follows:	
\begin{align*}
    p_{d,h} & = \sum_{i\in \left\lbrace 1,2,7\right\rbrace }\sum_{h^*=1}^{24}\b{i,h^*}^p p_{d-i,h^*} + \sum_{i=1}^3 \left( \b{i} p^{max}_{d-i} + \b{i+3} p^{min}_{d-i} \right) + \\
     & + \sum_{i=0}^1\sum_{h^*=1}^{24} \left( \b{i,h^*}^L l_{d-i,h^*} + \b{i,h^*}^S s_{d-i,h^*} + \b{i,h^*}^{on} w^{on}_{d-i,h^*} + \b{i,h^*}^{off} w^{off}_{d-i,h^*} \right) + \\
 & + \sum_{i=1}^7\b{6+i} D^i_{d} + \varepsilon_{d,h}, 
	\end{align*}
 
The ARX and fARX models are considered as baseline models that are estimated with regularization procedures discussed in the following Section \ref{ssec:regularization}.

\subsection{Regularization}
\label{ssec:regularization}

The concept of regularization involves minimizing the objective function of the original model and adding a penalty to the model parameters. In this study, the objective function is the residual sum of squares (RSS), and the regularization is defined by the following equation:

\begin{equation}
\label{eq:reg}
	\boldsymbol \beta = \argmin \underbrace{\sum_{d,h} \left( p_{d,h} - \sum_{i=1}^N \beta_i x^i_{d,h}\right) ^2}_\text{RSS} + \underbrace{\sum_{i=1}^N g(\boldsymbol{\beta};\lambda, \cdot)}_\text{penalty function},
\end{equation}

where $p_{d,h}$ represents the dependent variable (it is the price series) and $x_{d,h}$ is a matrix of independent variables consisting of model inputs. Finally, the $g(\boldsymbol \beta)$ is the penalty function, also known as the regularization term, applied to the parameter vector $\boldsymbol \beta = \left\lbrace \beta_1, \beta_2, \ldots, \beta_N \right\rbrace$.

The most commonly used form of regularization involves defining the penalty function as an order norm $\ell^1 $ or $\ell^2 $, scaled by a tuning parameter $\lambda$. However, numerous alternative approaches exist, which have yet to be tested in real-world applications. To our best knowledge, only three out of the ten considered regularization types have been previously tested in the EPF context and only LASSO is commonly used in the literature. In this study, ten penalty functions are used to generate the electricity price forecasts and their performance is compared against the unregularized benchmarks.

It should be noted that regularization can be utilized to identify the most important variables in the model by selecting an appropriate penalty function. This procedure involves fitting the full model with all predictors using an algorithm that reduces the coefficients of less significant explanatory variables toward zero. One of the benefits of using automatic variable selection methods is their ability to handle a large number of explanatory variables, thus reducing the dependence on expert knowledge, which is often unverified \citep{uni:now:wer:16}.

To conduct the research, we used the Matlab toolbox of McIlhagga~\cite{mci:16}. The toolbox utilizes Fisher scoring over an active set with orthant projection \cite{par:has:07,sch:fun:ros:09}. It offers an automated selection of the tuning parameter $\lambda$ using information criteria or cross-validation. In this study we decided to compare Bayesian information criteria and cross-validation with 7 folds. Moreover, for penalty functions with an extra regularization parameter, the toolbox allows one to utilize CV to select the optimal value of both parameters simultaneously, making the model fully automated (this feature is unavailable for BIC).

\subsubsection{Adaptive LASSO}

The \emph{adaptive least absolute shrinkage and selection operator} (aLASSO) penalty function was first introduced by Zou~\cite{zou:06}. The adaptive weights in the model are used for penalizing different coefficients in the $\ell^q$ penalty. According to the author, the adaptive LASSO performs as well as if the true underlying model was given in advance. Although originally the adaptive LASSO model was defined only for the $\ell^1$ penalty, it was further generalized to allow for the $\ell^q$ norm penalty. The adaptive penalty function is defined as follows:

\begin{equation}
g(\boldsymbol{\beta};\lambda,q) = \lambda  \frac{\left|\beta_i\right|}{\left|\beta^*_i\right|}^q,
\end{equation}

where $\beta^*_i$ is a vector of weights estimated with an unregularized model, $\lambda$ is the tuning parameter and $q$ is the additional parameter that refers to the $\ell^q$ norm. In this paper $\beta^*_i$ is defined as a vector of weights estimated with OLS, $\lambda$ is selected automatically via BIC or CV and $q$ takes one of the values from the grid $Q = \lbrace 1, 1.5, 2\rbrace$.

\subsubsection{Clipped LASSO}

The \emph{clipped least absolute shrinkage and selection operator} (cLASSO) \cite{ant:fan:01} is another penalty function that aims to generalize LASSO.
The idea of clipped LASSO is to set a threshold for the maximum value of the penalty for estimated weights. It is defined as follows:

\begin{equation}
g(\boldsymbol{\beta};\lambda,\alpha) = \lambda  \min \left\lbrace  \left|\beta_i\right|, \alpha \right\rbrace,
\end{equation}
where $\lambda$ is the tuning parameter and $\alpha$ is responsible for the threshold of the penalty. Based on limited pretests we consider $\alpha$ from a grid $A = \lbrace 0.5, 1, 1.5\rbrace$, whereas $\lambda$ is selected automatically via BIC or cross-validation.

\subsubsection{Concave potential function}

The regularization term of \emph{concave potential function} (CPF) \cite{nik:00} is the sum of the values obtained by applying the \emph{potential function} (PF) to each regressor. The penalty function is defined as follows:

\begin{equation}
g(\boldsymbol{\beta};\lambda,k) = \lambda  \frac{k\left|\beta_i\right|}{k+\left|\beta_i\right|},
\end{equation}

where $\lambda$ is the tuning parameter and $k$ is an additional parameter that refers to the shape of the penalty. Note that if $k \rightarrow \infty$ the penalty becomes equivalent to a standard LASSO penalty. In this paper, the additional parameter $k$ is selected from the initial grid $K = \lbrace 5, 15, 25\rbrace$.

\subsubsection{Elastic Net}

The \emph{elastic net} (EN) \cite{zou:has:05} can be viewed as an extension of ridge regression and LASSO. This penalty function incorporates a combination of linear and quadratic penalty terms:
\begin{equation}
g(\boldsymbol{\beta};\lambda,\alpha) = \lambda  (\alpha \left|\beta_i\right|+ (1-\alpha) \left|\beta_i\right|^2)
\end{equation}
where $\alpha \in [0,1]$. When $\alpha = 1$, the elastic net reduces to LASSO, and with $\alpha = 0$, it becomes a ridge regression. Note also that every elastic net problem can be rewritten as a LASSO problem on augmented data. Therefore, for fixed $\lambda$ and $\alpha$, the computational difficulty of the elastic net solution is similar to the LASSO problem \cite{has:tib:wai:15}. Just like in the other cases the value of $\lambda$ is automatically selected via BIC or cross-validation. The additional parameter $\alpha$ takes one of the values from the grid $A = \lbrace 0.25, 0.5, 0.75\rbrace$.

The elastic net was already applied in the context of forecasting electricity prices. In particular, it was first used by Uniejewski~et~al.~\cite{uni:now:wer:16}, where it produced the most accurate forecast across all considered automated variable selection methods. Recently, the method is gaining in popularity and more and more authors decide to estimate their forecasting models with elastic net \citep[for example][]{agr:muc:tri:19, mun:zie:20, nar:zie:20, cia:mun:zar:22}

\subsubsection{FLASH}
The \emph{forward-LASSO adaptive shrinkage} (FLASH)  was introduced by Radchenko and James~\cite{rad:jam:11}. To be precise, the FLASH algorithm does not fall specifically under penalized optimization techniques, but it has an implicit penalty. It follows a hierarchical approach similar to \emph{least-angle regression} \cite{efr:has:joh:tib:04} (LARS) and forward selection by incrementally adding one variable to the model, but with a unique feature of adjusting the level of shrinkage at each step to optimize the selection of the next variable. The penalty for coefficients in the active set is specified as:

\begin{equation}
	g(\boldsymbol{\beta};\lambda,\gamma) = \lambda  (1-\gamma)\left|\beta_i\right|,
\end{equation}

In this paper, the additional parameter $\gamma$ takes one of the values from the grid $G = \lbrace 0.25, 0.5, 0.75\rbrace$.

\subsubsection{LASSO}

\emph{Least absolute shrinkage and selection operator} (LASSO) was formally introduced by Tibshirani~\cite{tib:96} It has achieved great success in statistics and is widely used in various applications. The penalty function is defined as a $\ell^1$ norm scaled by the tuning parameter $\lambda$:

	\begin{equation}
	g(\boldsymbol{\beta};\lambda) = \lambda  \left|\beta_i\right|, 
	\end{equation}
where $\lambda$ is the tuning parameter. It indicates how significant the variables have to be to remain in the final model. While for $\lambda = 0$ the method reduces to \textit{ordinary least squares} (OLS), as the parameter increases, more and more variables are considered irrelevant and eliminated from the final model. In this research LASSO is treated as fully-automated model as the the value of $\lambda$  parameter is selected automatically via BIC or cross-validation.

LASSO is a regularization method that has been widely used in EPF but only in the past decade. Some of the earliest examples of research that employed LASSO in the context of EPF include \cite{lud:feu:neu:15,zie:ste:hus:15,gai:gou:ned:16,zie:16:TPWRS,zie:16:CSDA,uni:now:wer:16}.
 
\subsubsection{LQ}

LQ regularization \citep{gra:hal:sch:08} is another type of penalty function, after elastic net, that tries to fit between LASSO and ridge regression. This time instead of mixing linear and quadratic terms, the LQ regularization proposes to use a $\ell^q$ norm, for $q \in (1,2)$:

	\begin{equation}
	g(\boldsymbol{\beta};\lambda,q) = \lambda  \left|\beta_i\right|^q.
	\end{equation}
 
In this paper, we consider three values of $q$ from the grid $Q = \lbrace 1.25, 1.5, 1.75\rbrace$. Note that due to the toolbox requirements \cite{mci:16}, we had to impose the upper limit of the set of possible $\lambda$ values. Thus, in the LQ regularization, the tuning parameter cannot exceed a threshold equal to $2$. 
 
\subsubsection{MC+}

MC+ is a penalized variable selection method for high-dimensional linear regression \citep{zha:10, maz:fri:has:11}. Compared to LASSO, which is fast and continuous but biased, MC+ is nearly unbiased and accurate. The method consists of two elements: a \emph{minimax concave penalty} (MCP) and a \emph{penalized linear unbiased selection} (PLUS) algorithm. The MCP ensures the convexity of the penalized loss in sparse regions to the greatest extent possible, given certain thresholds for variable selection. The PLUS algorithm computes multiple exact local minimizers of a possibly nonconvex penalized loss function in a certain main branch of the graph of critical points of the penalized loss. The penalty is defined as follows:
	
	\begin{equation}
		g(\boldsymbol{\beta};\lambda,\gamma) = \left\{ \begin{array}{ll}
		\lambda\left|\beta_i\right| - \frac{\beta_i^2}{2\gamma}& \textrm{for $ \left|\beta_i\right| \leq \gamma\lambda$},\\
		\frac{\lambda^2\gamma}{2}& \textrm{for $\gamma\lambda < \left|\beta_i\right|$}.
	\end{array} \right. 
	\end{equation}
 
The additional parameter $\gamma\geq1$ takes one of the values from the grid $G = \lbrace 1, 3, 5\rbrace$. As for all other methods, the tuning parameter $\lambda$ is selected automatically via BIC or cross-validation. 
 
\subsubsection{Ridge}

Ridge regression is one of the first regularization method introduced in statistics \cite{hoe:ken:70}. Ridge regression is very similar to LASSO, but the linear penalty function is substituted with a quadratic norm:

	\begin{equation}
	g(\boldsymbol{\beta};\lambda) = \lambda  \left|\beta_i\right|^2
	\end{equation}
 
where $\lambda \ge 0$. Note that for $\lambda=0$, we get the standard OLS estimator; for $\lambda \rightarrow \infty$, all $\beta_{i}$'s tend to zero; while for intermediate values of $\lambda$, we are balancing two ideas: minimizing the RSS and shrinking the coefficients toward zero (and each other). The toolbox \cite{mci:16} requires us to specify the maximum value of $\lambda$. In this study, the maximum value has been set to 5.

Although ridge regression is one of the most classical forms of regularization, it very rarely appears in the EPF literature. Apart from Barnes~and~Balda~\cite{bar:bal:13} -- the first paper that utilizes ridge regression in the context of evaluating the profitability of battery storage -- the method has only been mentioned a couple of times \citep[for example][]{uni:now:wer:16, mir:mey:koc:17}.

\subsubsection{SCAD}

The smoothly clipped absolute deviation (SCAD) \cite{fan:li:01}  corresponds to a quadratic spline function with knots at $\lambda$ and $\alpha\lambda$. This penalty function leaves large values of $\beta$ not excessively penalized and makes the solution continuous. The resulting solution is given by:

	\begin{equation}
	g(\boldsymbol{\beta};\lambda,\alpha) =  \left\{ \begin{array}{ll}
	\lambda\left| \beta_i\right|  & \textrm{for $\left|\beta_i\right| \leq \lambda$},\\
	\frac{-\left| \beta_i\right| ^2+2\alpha\lambda\left|\beta_i\right| - \lambda^2}{2(\alpha-1)}& \textrm{for $\lambda<\left|\beta_i\right|| \leq \alpha \lambda$},\\
	\frac{\lambda^2(1+\alpha)}{2} & \textrm{for $ \lambda < \left|\beta_i\right|$},
	\end{array} \right. 
	\end{equation}

The additional parameter $\alpha\geq2$ takes one of the values from the grid $A = \lbrace 10, 20, 30\rbrace$.

\subsubsection{Penalties comparison}
\label{ssec:PenComp}

In this study, the considered penalties can be classified into three groups, each representing a different approach to regularization. The first group includes standard methods widely used in the literature, represented by LASSO and Ridge regression. These methods are based on simple $\ell^1$ or $\ell^2$ norms. Ridge regression uses a quadratic shrinkage factor that shrinks all $\beta_i$ parameters toward zero, but not exactly to zero, making it suitable for models with a small number of inputs. Conversely, LASSO focuses on both parameter shrinkage and variable selection, making it well suited for parameter-rich models.

\begin{figure}[t]
	\centering
	
	\includegraphics[width=0.45\textwidth]{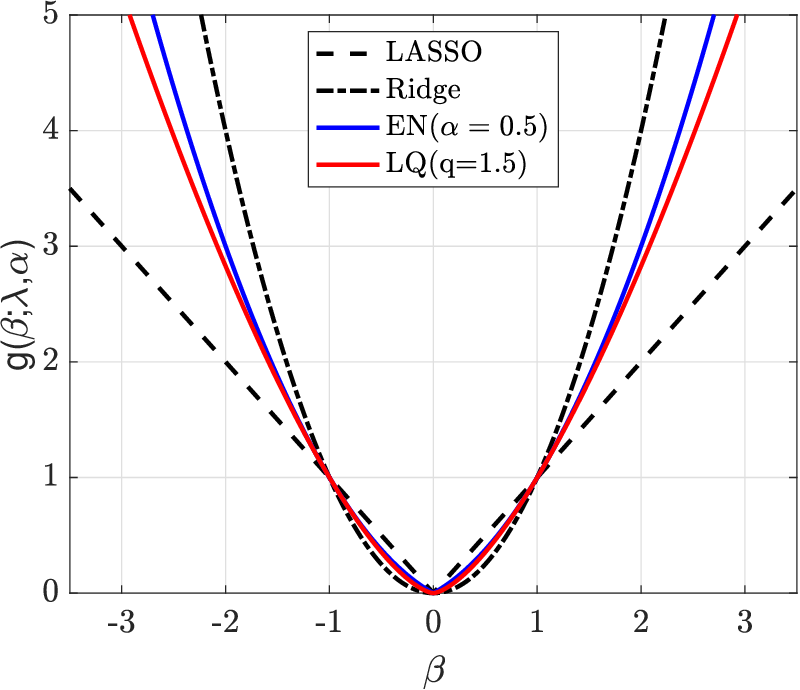} ~~
	\includegraphics[width=0.45\textwidth]{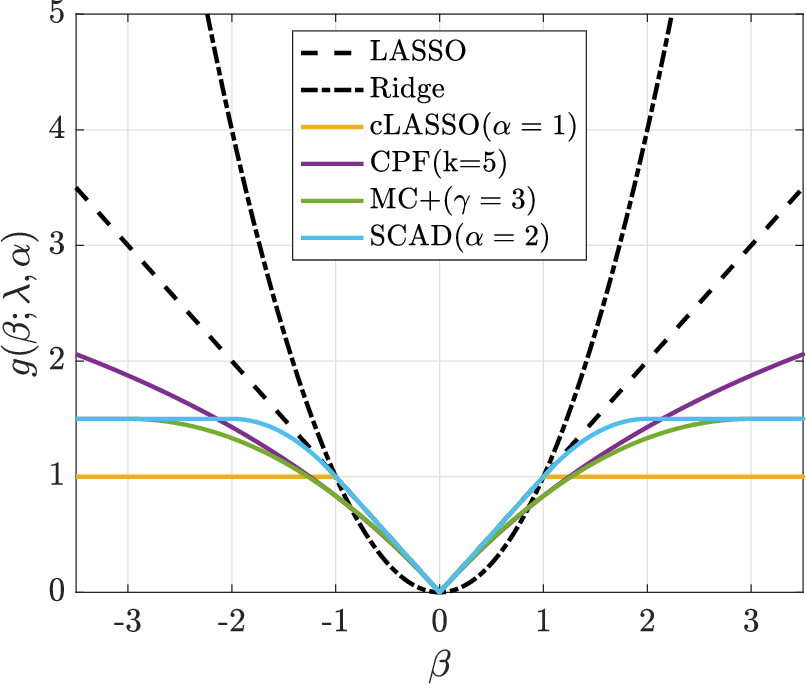}

	\caption{Shapes of different penalty functions (for $\lambda=1$) of the considered regularization types depending on the value of the estimated weight ($\beta$). For clarity, the plot is divided into two panels. The left panel shows penalty functions for the Elastic Net and LQ regularizations compared to standard approaches such as LASSO and Ridge regression. The right panel displays the penalties for all concave functions considered. Note that Adaptive LASSO and FLASH regularization are not included in the plot (see Section \ref{ssec:PenComp}).}
	\label{fig:reg}
\end{figure}

The second group includes methods that attempt to combine or enhance the capabilities of the standard approaches. These include Adaptive LASSO, Elastic Net, FLASH, and LQ regularization. The comparison between them is shown in the left panel of Figure \ref{fig:reg}. Note that Adaptive LASSO and FLASH regularization are not included in the plot. Adaptive LASSO, which depends on the $q$ parameter, has the same shape as LASSO, LQ, or Ridge, and differs only in the scale, which depends on the value of~$\beta^*$. FLASH, on the other hand, is not defined as a penalty, but rather as an optimization algorithm, so it cannot be easily plotted. 

The third group goes in a different direction and is represented by methods such as Clipped LASSO, Concave PF, MC+ or SCAD. These approaches use concave penalties, which means that the growth rate of the penalties decreases as the absolute value of $\beta$ increases. In particular, for Clipped LASSO, MC+, and SCAD, the penalty function becomes constant after a certain threshold. This allows the weights corresponding to less important variables to shrink to zero very quickly, while not penalizing $\beta_i$ too much for the most important inputs. The right panel of figure \ref{fig:reg} illustrates the comparison between these concave penalties, giving an insight into their characteristics.

\section{Results}
\label{sec:results}

Linear measures such as mean absolute error provide a simple and easy-to-understand assessment of the average forecasting error. However, linear measures have limitations, such as the inability to capture the variance of errors, which can lead to misleading conclusions. Therefore, in this study, we are using the \emph{root mean square errors} (RMSE). The measure is defined by the following equation:

\begin{equation}
\label{eqn:RMSE}
\text{RMSE} = \sqrt{\frac{1}{24D}\sum_{d=1}^{D} \sum_{h=1}^{24} \hat{\varepsilon}_{d,h}^2}.
\end{equation}

where, $\hat{\varepsilon}_{d,h} = {P_{d,h} -\hat{P}_{d,h}}$ is the forecasting error for day $d$ and hour $h$.

\subsection{OLS}

\begin{table}[t]
\caption{Root mean square error (RMSE) of the forecast obtained with two considered baseline models estimated with OLS. Bold text highlights the best model within each market.}
\label{tab:res:OLS}
\begin{tabular}{clcl||clcl}
\multicolumn{4}{c||}{EPEX}                              & \multicolumn{4}{c}{OMIE}                             \\
\multicolumn{2}{c|}{ARX}   & \multicolumn{2}{c||}{fARX}  & \multicolumn{2}{c|}{ARX}  & \multicolumn{2}{c}{fARX}  \\
\multicolumn{2}{c|}{13.55} & \multicolumn{2}{c||}{\textbf{12.65}} & \multicolumn{2}{c|}{\textbf{9.78}} & \multicolumn{2}{c}{10.35}
\end{tabular}
\end{table}

Table \ref{tab:res:OLS} shows the results of the ARX and fARX models estimated using the OLS method. It can be seen that, surprisingly, the less complex ARX model produces more accurate results for the OMIE market in terms of RMSE, and only slightly worse for the EPEX market. The relatively poor performance of the parameter-rich model is due to the overfitting problem.

The result of the OLS-estimated models presented in Table \ref{tab:res:OLS} should be considered as a reference point for other regulization-based estimation techniques. 

\subsection{Regularization results}
\subsubsection{Adaptive}
\label{ssec:res:Adaptive}
\begin{table}[b!]
    \caption{Root mean square error (RMSE) of the prediction obtained with two considered baseline models estimated with adaptive LASSO. The first three rows report results for models calibrated with a constant value of the parameter $q$ and cross-validation (CV) or Bayesian information criteria (BIC) to select the optimal tuning parameter $\lambda$. The last row corresponds to the model using cross-validation (CV) to select the optimal values of $q$ and $\lambda$. The best result within each model is underlined, while the best model within each market is additionally highlighted in bold.}
    \label{tab:res:Adaptive}
	\scalebox{1}{
  \begin{tabular}{r||cc|cc||cc|cc}
		& \multicolumn{4}{c||}{EPEX}  & \multicolumn{4}{c}{OMIE}                                                          \\[3pt]
		& \multicolumn{2}{c|}{ARX} & \multicolumn{2}{c||}{fARX} & \multicolumn{2}{c|}{ARX} & \multicolumn{2}{c}{fARX}  \\[3pt]
		& CV         & BIC        & CV         & BIC         & CV             & BIC            & CV         & BIC \\
  \midrule
        1   & {\ul 14.61} & 15.67 & {\ul \textbf{12.12}} & 15.39 & {\ul 9.79} & 10.20 & 9.63       & 10.26 \\
1.5 & 15.06       & 15.84 & 12.13       & 15.32 & 9.81       & 10.26 & 9.62       & 10.27 \\
2   & 15.28       & 15.78 & 12.19       & 15.42 & 9.84       & 10.28 & {\ul \textbf{9.58}} & 10.33 \\
CV  & 14.68       &       & 12.17       &       & 9.79       &       & 9.62       &      
	\end{tabular}
}
\end{table}

Table \ref{tab:res:Adaptive} shows the RMSE of the ARX and fARX models estimated using the adaptive LASSO method. The results for each market are divided into two sub-columns referring to two model structures (ARX and fARX). In addition, for each model, the results are reported for two $\lambda$ selection methods. Finally, the first three rows refer to a model with a fixed $q$ parameter, and the last row (CV) presents the result of the fully automated approach when both $\lambda$ and $q$ are selected by cross-validation. Some important conclusions can be drawn from the results presented in the Table \ref{tab:res:Adaptive}:
\begin{itemize}
    \item The parameter-rich fARX model outperforms the parsimonious ARX model in both markets, although the difference is more pronounced in the EPEX market.
    \item Cross-validation is a better method in terms of prediction accuracy for selecting the $\lambda$ parameter than BIC for adaptive LASSO, especially when using the parameter-rich fARX model.
    \item Among all considered values of the additional parameter, $q = 1$ performs best on average in terms of RMSE. However, the fully automated approach is only slightly worse (by less than 0.5\%) than the optimal value of the parameter. This indicates that CV can be successfully applied to select values of both parameters ($\lambda$ and $q$).
\end{itemize}

\subsubsection{Clipped LASSO}
\label{ssec:res:Clipped}
\begin{table}[t]
    \caption{Root mean square error (RMSE) of the prediction obtained with two considered baseline models estimated with clipped LASSO. The first three rows report results for models calibrated with a constant value of the parameter $\alpha$ and cross-validation (CV) or Bayesian information criteria (BIC) to select the optimal tuning parameter $\lambda$. The last row corresponds to the model using cross-validation (CV) to select the optimal values of $\alpha$ and $\lambda$. The best result within each model is underlined, while the best model within each market is additionally highlighted in bold.}
    \label{tab:res:Clipped}
	\scalebox{1}{
  \begin{tabular}{r||cc|cc||cc|cc}
		& \multicolumn{4}{c||}{EPEX}  & \multicolumn{4}{c}{OMIE}                                                          \\[3pt]
		& \multicolumn{2}{c|}{ARX} & \multicolumn{2}{c||}{fARX} & \multicolumn{2}{c|}{ARX} & \multicolumn{2}{c}{fARX}  \\[3pt]
		& CV         & BIC        & CV         & BIC         & CV             & BIC            & CV         & BIC \\
  \midrule
0.5 & 13.93       & 15.28 & 11.96       & 17.56 & {\ul 9.74} & 10.18 & 9.39       & 12.03 \\
1   & {\ul 13.91} & 15.20 & 11.95       & 17.48 & 9.80       & 10.53 & 9.39       & 12.17 \\
1.5 & 13.91       & 15.23 & 11.95       & 17.45 & 9.81       & 10.58 & {\ul \textbf{9.38}} & 12.25 \\
CV  & 13.91       &       & {\ul \textbf{11.95}} &       & 9.77       &       & 9.39       &       
	\end{tabular}
}
\end{table}

Table \ref{tab:res:Clipped} reports the RMSE of the models estimated using the clipped LASSO method. The results are presented in a format similar to the tables in the previous sections. In particular, the results for each market are reported for two model structures (ARX and fARX) and two methods of $\alpha$ selection. There are several important conclusions that can be drawn from the results presented in the Table \ref{tab:res:Clipped}:
\begin{itemize}
    \item Similar to the adaptive LASSO, the fARX model outperforms the parsimonious ARX model.
    \item Considering the method used to select the $\lambda$ parameter, the CV beats the BIC by a large margin.
    \item The optimal value of the parameter $\gamma$ varies across models and markets. This time the fully automated approach performs best for EPEX market and only slightly worse (for OMIE) than the best \emph{ex-post} value of the parameter.
\end{itemize}

\subsubsection{Concave PF}
\label{ssec:res:Concave}
\begin{table}[b!]
    \caption{Root mean square error (RMSE) of the prediction obtained with two considered baseline models estimated with concave PF. The first three rows report results for models calibrated with a constant value of the parameter $k$ and cross-validation (CV) or Bayesian information criteria (BIC) to select the optimal tuning parameter $\lambda$. The last row corresponds to the model using cross-validation (CV) to select the optimal values of $k$ and $\lambda$. The best result within each model is underlined, while the best model within each market is additionally highlighted in bold.}
    \label{tab:res:Concave}
	\scalebox{1}{
  \begin{tabular}{r||cc|cc||cc|cc}
& \multicolumn{4}{c||}{EPEX}  & \multicolumn{4}{c}{OMIE}                                                          \\[3pt]
		& \multicolumn{2}{c|}{ARX} & \multicolumn{2}{c||}{fARX} & \multicolumn{2}{c|}{ARX} & \multicolumn{2}{c}{fARX}  \\[3pt]
		& CV         & BIC        & CV         & BIC         & CV             & BIC            & CV         & BIC \\
  \midrule
        5  & 13.93       & 15.37 & 11.97       & 17.67 & 9.81       & 10.46 & 9.42       & 11.87 \\
15 & 13.92       & 15.22 & 11.96       & 17.59 & {\ul 9.81} & 10.51 & {\ul \textbf{9.41}} & 11.99 \\
25 & {\ul 13.91} & 15.22 & {\ul \textbf{11.96}} & 17.54 & 9.81       & 10.54 & 9.41       & 11.92 \\
CV & 13.93       &       & 11.96       &       & 9.82       &       & 9.41       &      
	\end{tabular}
}
\end{table}

Table \ref{tab:res:Concave} presents the RMSE of the forecasts obtained with models estimated using the concave PF regularization method.  Some important conclusions can be drawn from the results presented in Table \ref{tab:res:Concave}:
\begin{itemize}
    \item As previously, the most accurate forecasts are produced by the parameter-rich model (fARX). 
    \item In terms of selecting the value of the $\lambda$ parameter, cross-validation clearly outperforms BIC, especially in the case of the fARX baseline model where BIC struggles to select the optimal $\lambda$ value.
    \item The value $k=25$ performs best on average in terms of predictive accuracy, outperforming all other considered values of the additional parameter. The fully automated approach shows only a small performance loss compared to the optimal value of the parameter (less than 0.1\%), indicating that it is a good performing alternative.
\end{itemize}

\subsubsection{Elastic Net}
\label{ssec:res:Elastic}
\begin{table}[t]
    \caption{Root mean square error (RMSE) of the prediction obtained with two considered baseline models estimated with elastic net. The first three rows report results for models calibrated with a constant value of the parameter $\alpha$ and cross-validation (CV) or Bayesian information criteria (BIC) to select the optimal tuning parameter $\lambda$. The last row corresponds to the model using cross-validation (CV) to select the optimal values of $\alpha$ and $\lambda$. The best result within each model is underlined, while the best model within each market is additionally highlighted in bold.}
    \label{tab:res:Elastic}
	\scalebox{1}{
  \begin{tabular}{r||cc|cc||cc|cc}
		& \multicolumn{4}{c||}{EPEX}  & \multicolumn{4}{c}{OMIE}                                                          \\[3pt]
		& \multicolumn{2}{c|}{ARX} & \multicolumn{2}{c||}{fARX} & \multicolumn{2}{c|}{ARX} & \multicolumn{2}{c}{fARX}  \\[3pt]
		& CV         & BIC        & CV         & BIC         & CV             & BIC            & CV         & BIC \\
  \midrule
        0.25 & {\ul 13.81} & 14.94 & {\ul \textbf{11.92}} & 20.33 & 9.81       & 10.55 & 9.35       & 12.62 \\
0.5  & 13.84       & 14.87 & 11.95       & 18.35 & 9.80       & 10.61 & 9.34       & 12.03 \\
0.75 & 13.87       & 15.02 & 11.96       & 18.16 & {\ul 9.79} & 10.48 & {\ul \textbf{9.31}} & 11.76 \\
CV   & 13.87       &       & 11.96       &       & 9.79       &       & 9.32       &      
	\end{tabular}
}
\end{table}

Table \ref{tab:res:Elastic} displays the RMSE of the elastic net estimated models. The results for each market are presented in the same table format as in the previous sections. The results presented in the Table \ref{tab:res:Elastic} allow us to draw some important conclusions:
\begin{itemize}
    \item The fARX model produces the most accurate forecasts. However, the ARX model performs better when the value of the $\lambda$ parameter is selected with BIC.
    \item Cross-validation clearly outperforms BIC, with the difference being more visible for fARX model.
    \item The values $\alpha=0.25$ and $\alpha=0.75$ perform best for EPEX and OMIE, respectively. The fully automated approach tends to produce very similar results (in terms of prediction accuracy) to the $\alpha=0.75$ model.
\end{itemize}
\subsubsection{FLASH}
\label{ssec:res:Flash}
\begin{table}[b!]
    \caption{Root mean square error (RMSE) of the prediction obtained with two considered baseline models estimated with FLASH algorithm. The first three rows report results for models calibrated with a constant value of the parameter $\gamma$ and cross-validation (CV) or Bayesian information criteria (BIC) to select the optimal tuning parameter $\lambda$. The last row corresponds to the model using cross-validation (CV) to select the optimal values of $\gamma$ and $\lambda$. The best result within each model is underlined, while the best model within each market is additionally highlighted in bold.}
    \label{tab:res:flash}
	\scalebox{1}{
  \begin{tabular}{r||cc|cc||cc|cc}
		& \multicolumn{4}{c||}{EPEX}  & \multicolumn{4}{c}{OMIE}                                                          \\[3pt]
		& \multicolumn{2}{c|}{ARX} & \multicolumn{2}{c||}{fARX} & \multicolumn{2}{c|}{ARX} & \multicolumn{2}{c}{fARX}  \\[3pt]
		& CV         & BIC        & CV         & BIC         & CV             & BIC            & CV         & BIC \\
  \midrule
        0.25 & 13.92       & 15.35 & {\ul \textbf{11.97}} & 15.73 & {\ul 9.83} & 10.52 & {\ul \textbf{9.40}} & 10.69 \\
0.5  & 13.90       & 15.33 & 11.98       & 15.21 & 9.85       & 10.45 & 9.41       & 10.53 \\
0.75 & {\ul 13.81} & 15.29 & 11.97       & 14.61 & 9.92       & 10.41 & 9.46       & 10.32 \\
CV   & 13.82       &       & 11.98       &       & 9.92       &       & 9.42       &  
	\end{tabular}
}
\end{table}

Table \ref{tab:res:flash} shows the RMSE of the forecasts obtained with models estimated using the FLASH regularization. The results are presented in a format similar to the tables in the previous sections. In particular, for each market, the results are divided into two model structures (ARX and fARX) and two methods of $\lambda$ selection. From the results in Table \ref{tab:res:flash}, some important conclusions can be drawn:
\begin{itemize}
    \item The fARX clearly outperforms the ARX model.
    \item Regarding the choice of the $\lambda$ parameter, cross-validation outperforms BIC by a substantial margin, especially for the EPEX market.
    \item The value $\gamma=0.25$ performs best on average for both markets. The fully automated approach tends to produce forecasts that are only slightly less accurate than the best model.
\end{itemize}

\subsubsection{LASSO}
\label{ssec:res:LASSO}
\begin{table}[t]
    \caption{Root mean square error (RMSE) of the forecast obtained with two considered baseline models estimated with LASSO regularization. The best result within each model is underlined, while the best model within each market is additionally highlighted in bold.}
    \label{tab:res:LASSO}
\begin{tabular}{cc|cc||cc|cc}
\multicolumn{4}{c||}{EPEX}                           & \multicolumn{4}{c}{OMIE}                           \\
\multicolumn{2}{c}{ARX} & \multicolumn{2}{c||}{fARX} & \multicolumn{2}{c}{ARX} & \multicolumn{2}{c}{fARX} \\
CV            & BIC     & CV             & BIC     & CV            & BIC     & CV            & BIC      \\
\midrule
{\ul 13.91}   & 15.23   & {\ul \textbf{11.94}}    & 17.74   & {\ul 9.81}    & 10.57   & {\ul \textbf{9.39}}    & 12.63   
\end{tabular}
\end{table}

The RMSE of the models estimated using LASSO are presented in table \ref{tab:res:LASSO}. The results are divided into two model structures (ARX and fARX) and two methods of selecting $\lambda$ values for each market. Note that the LASSO regularization has only one parameter, so both CV and BIC are fully automated approaches. The results in Table \ref{tab:res:LASSO} lead to several important conclusions:
\begin{itemize}
    \item The fARX model produces the most accurate price predictions, but ARX outperforms fARX in some cases.
\item Cross-validation beats BIC by a large margin for selecting the $\lambda$ parameter, especially for the fARX model.
\end{itemize}

\subsubsection{Lq}
\label{ssec:res:Lq}
\begin{table}[b!]
    \caption{Root mean square error (RMSE) of the prediction obtained with two considered baseline models estimated with adaptive LASSO. The first three rows report results for models calibrated with a constant value of the parameter $q$ and cross-validation (CV) or Bayesian information criteria (BIC) to select the optimal tuning parameter $\lambda$. The last row corresponds to the model using cross-validation (CV) to select the optimal values of $q$ and $\lambda$. The best result within each model is underlined, while the best model within each market is additionally highlighted in bold.}
    \label{tab:res:Lq}
	\scalebox{1}{
  \begin{tabular}{r||cc|cc||cc|cc}
		& \multicolumn{4}{c||}{EPEX}  & \multicolumn{4}{c}{OMIE}                                                          \\[3pt]
		& \multicolumn{2}{c|}{ARX} & \multicolumn{2}{c||}{fARX} & \multicolumn{2}{c|}{ARX} & \multicolumn{2}{c}{fARX}  \\[3pt]
		& CV         & BIC        & CV         & BIC         & CV             & BIC            & CV         & BIC \\
  \midrule
        1.25 & 13.85 & 13.54       & {\ul \textbf{11.92}} & 32.68 & 9.79 & {\ul 9.77} & {\ul \textbf{9.33}} & 28.15 \\
1.5  & 13.79 & 13.54       & 11.92       & 29.42 & 9.80 & 9.77       & 9.34       & 22.40 \\
1.75 & 13.74 & {\ul 13.54} & 11.95       & 22.70 & 9.82 & 9.77       & 9.39       & 16.13 \\
CV   & 13.85 &             & 11.93       &       & 9.78 &            & 9.34       &      
	\end{tabular}
}
\end{table}

The RMSE of the predictions estimated with LQ regularization is shown in the table \ref{tab:res:Lq}. The presentation of the results follows a similar format to the previous sections, where the results are divided into two model structures (ARX and fARX) and two methods of selecting the value of the $\lambda$ parameter for each market. The first three rows in each panel correspond to a model with a fixed parameter $\alpha$, while the last row (CV) represents the results of the fully automated approach.. From the results in Table \ref{tab:res:Lq}, several important conclusions can be drawn:
\begin{itemize}
    \item The fARX model produces the best prediction in terms of forecast accuracy. However, it performs very weakly when BIC is used to select the value of the $\lambda$ parameter.
\item The performance of the two methods for selecting the optimal value of the $\lambda$ parameter depends on the structure of the model. In particular, BIC performs better for the parsimonious ARX model, but fails to select $\lambda$ for the more complex fARX model.
    \item The value $q=1.25$ performs best on average in terms of RMSE. The fully automated approach performs only slightly worse than the optimal \textit{ex-post} model for the fARX model, but substantially worse for the ARX structure.
\end{itemize}

\newpage
\subsubsection{MC+}
\label{ssec:res:Mcp}
\begin{table}[t]
    \caption{Root mean square error (RMSE) of the prediction obtained with two considered baseline models estimated with MC+. The first three rows report results for models calibrated with a constant value of the parameter $\gamma$ and cross-validation (CV) or Bayesian information criteria (BIC) to select the optimal tuning parameter $\lambda$. The last row corresponds to the model using cross-validation (CV) to select the optimal values of $\gamma$ and $\lambda$. The best result within each model is underlined, while the best model within each market is additionally highlighted in bold.}
    \label{tab:res:Mcp}
	\scalebox{1}{
  \begin{tabular}{r||cc|cc||cc|cc}
		& \multicolumn{4}{c||}{EPEX}  & \multicolumn{4}{c}{OMIE}                                                          \\[3pt]
		& \multicolumn{2}{c|}{ARX} & \multicolumn{2}{c||}{fARX} & \multicolumn{2}{c|}{ARX} & \multicolumn{2}{c}{fARX}  \\[3pt]
		& CV         & BIC        & CV         & BIC         & CV             & BIC            & CV         & BIC \\
  \midrule
1  & 13.98       & 15.59 & 12.00       & 16.27 & 9.83       & 10.16 & 9.41       & 10.52 \\
3  & 13.93       & 15.35 & 11.99       & 16.98 & {\ul 9.81} & 10.47 & 9.41       & 11.21 \\
5  & {\ul 13.92} & 15.27 & {\ul \textbf{11.98}} & 17.11 & 9.81       & 10.45 & {\ul \textbf{9.40}} & 11.36 \\
CV & 13.99       &       & 12.00       &       & 9.84       &       & 9.40       &      

	\end{tabular}
}
\end{table}

Table \ref{tab:res:Mcp} shows the RMSE of the models estimated by the MC+ regularization. The format of the table is the same as in the previous sections. There are some important conclusions that can be drawn from the results in the Table \ref{tab:res:Mcp}:
\begin{itemize}
    \item The fARX model outperforms the ARX model on average for both methods of selecting the tuning parameter $\lambda$.
    \item Cross-validation consistently outperforms BIC for $\lambda$ value selection by a large margin, especially in the EPEX market.
    \item The optimal value of the parameter $\gamma$ varies across models, but on average the $\gamma=5$ performs best. While the fully automated approach is slightly outperformed by the optimal parameter values in each case, it remains a reasonable alternative. This suggests that using a fully automated approach can lead to satisfying results without having to rely on expert parameter choices.
\end{itemize}

\subsubsection{Ridge}
\label{ssec:res:Ridge}
\begin{table}[b]
    \caption{Root mean square error (RMSE) of the forecast obtained with two considered baseline models estimated with ridge regularization. The best result within each model is underlined, while the best model within each market is additionally highlighted in bold.}
    \label{tab:res:Ridge}
\begin{tabular}{cc|cc||cc|cc}
\multicolumn{4}{c||}{EPEX}                           & \multicolumn{4}{c}{OMIE}                           \\
\multicolumn{2}{c}{ARX} & \multicolumn{2}{c||}{fARX} & \multicolumn{2}{c}{ARX} & \multicolumn{2}{c}{fARX} \\
CV            & BIC     & CV             & BIC     & CV            & BIC     & CV            & BIC      \\
\midrule
13.70 & {\ul 13.53} & {\ul \textbf{12.02}} & 22.79 & 9.83 & {\ul 9.77} & {\ul \textbf{9.39}} & 15.99 \\ 
\end{tabular}
\end{table}

In table \ref{tab:res:Ridge} we present the RMSE of the models estimated by ridge regression. The results are divided into two model structures (ARX and fARX) and two methods of selecting $\lambda$ values for each market. Ridge regression, similar to LASSO, has only one parameter, and thus both CV and BIC are fully automated approaches. There are several important conclusions that can be drawn from the results presented in the table \ref{tab:res:Ridge}:
\begin{itemize}
    \item The parameter-rich fARX model produces the most accurate forecasts, but in the ARX outperforms fARX.
    \item In terms of selecting the value of the parameter $\lambda$, CV outperforms the information criteria for the fARX model, but BIC performs better for the parsimonious ARX model.
\end{itemize}

\subsubsection{SCAD}
\label{ssec:res:Scad}
\begin{table}[t]
    \caption{Root mean square error (RMSE) of the prediction obtained with two considered baseline models estimated with SCAD regularization. The first three rows report results for models calibrated with a constant value of the parameter $\alpha$ and cross-validation (CV) or Bayesian information criteria (BIC) to select the optimal tuning parameter $\lambda$. The last row corresponds to the model using cross-validation (CV) to select the optimal values of $\alpha$ and $\lambda$. The best result within each model is underlined, while the best model within each market is additionally highlighted in bold.}
    \label{tab:res:Scad}
	\scalebox{1}{
  \begin{tabular}{r||cc|cc||cc|cc}
		& \multicolumn{4}{c||}{EPEX}  & \multicolumn{4}{c}{OMIE}                                                          \\[3pt]
		& \multicolumn{2}{c|}{ARX} & \multicolumn{2}{c||}{fARX} & \multicolumn{2}{c|}{ARX} & \multicolumn{2}{c}{fARX}  \\[3pt]
		& CV         & BIC        & CV         & BIC         & CV             & BIC            & CV         & BIC \\
  \midrule
        10 & {\ul 13.71} & 14.75 & 12.18       & 15.00 & 9.98       & 10.13 & {\ul \textbf{9.47}} & 10.10 \\
20 & 13.82       & 15.03 & 12.15       & 14.85 & 9.91       & 10.12 & 9.51       & 10.22 \\
30 & 13.86       & 14.91 & {\ul \textbf{12.13}} & 15.06 & {\ul 9.85} & 10.10 & 9.56       & 10.38 \\
CV & 13.76       &       & 12.15       &       & 9.97       &       & 9.51       &   
	\end{tabular}
}
\end{table}

Table \ref{tab:res:Scad} presents the RMSE of the models estimated using the SCAD regularization. The presentation of the results follows a similar format to the previous sections, where the results are divided into two model structures (ARX and fARX) and two methods of selecting the value of the $\lambda$ parameter for each market. The first three rows in each panel correspond to a model with a fixed $\alpha$ parameter, while the last row (CV) represents the results of the fully automated approach. Several important conclusions can be drawn from the results presented in the Table \ref{tab:res:Scad}:
\begin{itemize}
    \item The fARX model outperforms the ARX model on average for both methods of selecting the tuning parameter $\lambda$.
    \item For selecting the value of the $\lambda$ parameter, cross-validation consistently outperforms BIC by a large margin.
    \item It is impossible to identify a clear winner among models and markets with respect to the optimal value of the $\alpha$ parameter, as it varies among them. Despite the fact that the fully automated approach is slightly inferior to the optimal parameter values in most cases, it is still a valuable option. This means that relying on a fully automated approach can yield satisfying results.
\end{itemize}

\subsection{Summary of results}

Sections \ref{ssec:res:Adaptive}-\ref{ssec:res:Scad} presented the results for each regularization technique separately. While the conclusions varied depending on the penalty function used, there were some common factors that remained consistent across most regularization types. Based on the results in Tables \ref{tab:res:Adaptive}-\ref{tab:res:Scad}, the most important conclusions can be summarized as follows:

\begin{itemize}
\item In general, the fARX model with cross-validation as the $\lambda$ selection method produces the most accurate forecasts for both markets. Although BIC slightly outperforms CV in some cases, this is only true for the parsimonious ARX model, which on average produces less accurate forecasts compared to fARX.
\item Overall, BIC is not recommended for selecting the tuning parameter $\lambda$, especially for models with a large number of parameters.
\item For the vast majority of regularization types, the accuracy of the forecast obtained with the fully automated approach, where all parameters were selected by cross-validation, is very similar to the best model selected \textit{ex-post}. This shows that the presented solutions are robust and can be applied to different markets and time series without pre-selecting their parameters.
\end{itemize}

In the previous sections, the accuracy of the forecasts was only compared within a certain penalty function. However, the purpose of this study is to identify the best type of regularization in the context of electricity price forecasting. To rank the penalty functions, Table \ref{tab:res} summarizes the results for all ten considered regularization types. Table \ref{tab:res} shows the percentage change compared to the OLS estimated model. The relative measure is defined as
    $$
    rRMSE^{reg} = \frac{RMSE^{reg}-RMSE^{OLS}}{RMSE^{OLS}} \times 100\%
$$
where $RMSE^{reg}$ is the root mean squared error of the given regularization type. Meanwhile, $RMSE^{OLS}$ is the RMSE value for the predictions obtained with the corresponding OLS-estimated model.

\begin{table}[b!]
\caption{The relative root mean square errors of the forecasts obtained using different regularization techniques, compared to the models estimated using OLS. The results for each model are presented in two columns. The first column (CV) shows the result of the model with all parameters automatically selected by cross-validation. The second column (best) shows the result of the model with the $\lambda$ parameter selected by CV or BIC (the better one selected \emph{ex-post}) and the other parameter selected from the grid according to \textit{ex-post} performance. Note that the best result in each column is bolded and any result better than LASSO is underlined.}
\label{tab:res}
         \begin{tabular}{r||cc|cc||cc|cc}
		& \multicolumn{4}{c||}{EPEX}  & \multicolumn{4}{c}{OMIE}                                                          \\[3pt]
		& \multicolumn{2}{c|}{ARX} & \multicolumn{2}{c||}{fARX} & \multicolumn{2}{c|}{ARX} & \multicolumn{2}{c}{fARX}  \\[3pt]
		& CV         & best        & CV         & best         & CV             & best            & CV         & best \\
  \midrule
OLS      & \multicolumn{2}{c|}{-}        & \multicolumn{2}{c||}{-}         & \multicolumn{2}{c|}{-}        & \multicolumn{2}{c}{-}           \\
aLASSO & 8.33\%       & 7.88\%        & -3.83\%       & -4.19\%       & {\ul 0.13\%} & {\ul 0.10\%}  & -7.13\%        & -7.44\%        \\
cLASSO  & 2.69\% & {\ul 2.65\%} & -5.56\% & -5.56\% & {\ul 0.00\%} & {\ul \textbf{-0.36\%}} & {\ul -9.32\%} & {\ul -9.44\%} \\
CPF  & 2.82\%       & 2.69\%        & -5.47\%       & -5.49\%       & 0.41\%       & 0.37\%        & -9.10\%        & -9.15\%        \\
EN  & {\ul 2.41\%} & {\ul 1.97\%}  & -5.48\%       & {\ul -5.79\%} & {\ul 0.19\%} & {\ul 0.17\%}  & \textbf{{\ul -10.00\%}} & \textbf{{\ul -10.11\%}} \\
FLASH    & {\ul 2.00\%} & {\ul 1.97\%}  & -5.34\%       & -5.39\%       & 1.51\%       & 0.52\%        & -9.01\%        & -9.25\%        \\
LASSO    & \multicolumn{2}{c|}{2.66\%}        & \multicolumn{2}{c||}{-5.64\%}       &  \multicolumn{2}{c|}{0.32\%}        & \multicolumn{2}{c}{-9.29\%}        \\
LQ       & {\ul 2.26\%} & {\ul -0.03\%} & \textbf{{\ul -5.71\%}} & \textbf{{\ul -5.84\%}} & {\ul 0.09\%} & {\ul -0.06\%} & {\ul -9.80\%}  & {\ul -9.91\%}  \\
MC       & 3.27\% & 2.76\%       & -5.17\% & -5.31\% & 0.66\%                & 0.40\%                 & -9.21\%       & -9.23\%       \\
Ridge    & \multicolumn{2}{c|}{{\ul \textbf{-0.11\%}}} & \multicolumn{2}{c||}{-4.98\%}       & \multicolumn{2}{c|}{\textbf{{\ul -0.09\%}}} & \multicolumn{2}{c}{-9.29\%}        \\
SCAD     & {\ul 1.57\%} & {\ul 1.20\%}  & -4.01\%       & -4.12\%       & 1.94\%       & 0.72\%        & -8.16\%        & -8.53\%       
\end{tabular}
\end{table}

Table \ref{tab:res} shows the percentage change compared to OLS estimation for the fully automated approaches and the best \textit{ex-post} models. For each model and market, we present the result of the procedure when all parameters are selected with CV and the results underlined in Tables \ref{tab:res:Adaptive}-\ref{tab:res:Scad}. Note that the negative value in Table \ref{tab:res} indicates that the model estimated with a given regularization type is better in terms of predictive accuracy than the same model estimated with OLS. Moreover, considering that the most common regularization technique in the EPF application is LASSO, the results of models that outperform LASSO-estimated equivalents are underlined.

Several important conclusions can be drawn from the results reported in Table \ref{tab:res}:
\begin{itemize}
    \item The most accurate forecasts for the parsimonious ARX model are obtained with a fully automated approach using ridge regression and the BIC method to select the tuning parameter $\lambda$.
    \item Overall, the OLS estimation is hard to beat for the ARX model. Only two types of regularization managed to produce more accurate forecasts for both markets, and only by a small margin. In particular, the ARX model estimated with the LQ regularization with the $q$ parameter selected \emph{ex-post} outperformed the OLS estimation by 0.03\% and 0.06\% for the EPEX and OMIE markets, respectively. 
    \item On the other hand, for the parameter-rich fARX model, estimation with any regularization function outperforms the OLS-estimated model for both markets.
    \item In the case of the fARX model, the LASSO estimation is very hard to beat for other regularization functions. Only the elastic net and LQ are able to consistently outperform LASSO.
    \item Across all the regularization types, the LQ regularization performs best on average. It produces more accurate forecasts than the LASSO-estimated models in all cases.
    \item The second best regularization type in terms of forecast accuracy is the elastic net. In particular, it outperforms LASSO in seven out of eight cases. 
    \item It is worth noting the very good performance of Clipped LASSO for the OMIE market. It performs best of all regularization types for ARX and outperforms LASSO for fARX. However, it is outperformed by other methods for the EPEX market.
    \end{itemize}

\section{Conclusions}
\label{sec:conclusions}

The study investigated the use of different types of penalties for regularization in the context of electricity price forecasting. It compares the performance of ten regularization techniques, including adaptive LASSO, clipped LASSO, concave potential function, elastic net, forward LASSO adaptive shrinkage, LASSO, LQ, minimax concave PLUS, ridge regression, and smoothly clipped absolute deviation, using two data sets from the German EPEX SPOT and Iberian OMIE markets and two different model structures. 

The results show that for both markets, the parameter-rich fARX model with cross-validation as the parameter selection method produces the most accurate forecasts. Conversely, using BIC to select the optimal $\lambda$ value is not advisable for the majority of regularized models. The study recommends using cross-validation to select all parameters in a fully automated manner. The accuracy of such models is similar to the best models selected \textit{ex-post}, and at the same time it does not require additional expert knowledge.

OLS estimation is hard to beat for the parsimonious ARX model, but ridge regression slightly outperforms the benchmark. For the parameter-rich fARX model, the estimation with any regularization function far outperforms the OLS-estimated model for both markets. The analysis shows that the LQ and elastic net techniques perform better than others on average in terms of RMSE. In particular, they were the only types of penalty functions that consistently produced more accurate forecasts than the most commonly used LASSO. 

\section*{Acknowledgments}
This work was partially supported by the Ministry of Science and Higher Education (MNiSW, Poland) through Diamond Grant No. 0199/DIA/2019/48.

\bibliographystyle{acm}
\bibliography{references}

\end{document}